\documentclass[aps,preprint,amsmath,amssymb]{revtex4}

\usepackage{graphicx}
\begin{document}

\title{Joint analysis of solar neutrino and new KamLAND data in the RSFP framework } 

\author{D. Yilmaz}
\email[]{dyilmaz@wisc.edu} \email[]{dyilmaz@eng.ankara.edu.tr}
\affiliation{Department of Physics, University of Wisconsin,
Madison, WI 53706, USA}
\affiliation{Department of Physics Engineering, Faculty of Engineering, Ankara University, 06100 Tandogan, Ankara, TURKEY}

\begin{abstract}
A joint analysis of solar neutrino data together with the new KamLAND data is presented in the RSFP framework. It is investigated that how the new KamLAND data effects the allowed regions at different $\mu B$ values. A limit on $\mu B$ value is found at the different confidence level intervals. It is shown that the RSFP scenerio does not have a crucial role on the solar neutrino data. 
\end{abstract}

\maketitle

\section{INTRODUCTION}
After the solar neutrino experiments showed that solar neutrino fluxes reduced compared to the standard solar model predictions [1], as a possible mechanism, neutrino oscillations were proposed to explain this neutrino deficit. In addition to the vacuum oscillations, Mikheyev-Simirnov-Wolfenstein (MSW) [2-3] effect in the oscillation was proposed as another solution to the solar neutrino deficit. In MSW solution, when neutrino is passing through the matter, a resonant enhancement of neutrino oscillation appears. Earlier solar neutrino experiments, chlorine [4] and three galium measurements [5-7], and later Super-Kamiokande (SK) [8] and SNO [9-10] confirmed the neutrino oscillation and global analysis of them showed that the so-called large mixing angle (LMA) region of the neutrino parameter space was the most likely solution [11]. Neutrino oscillation is known as implication of the new physics beyond the Standard Model. In a minimal extension of the Standard Model, neutrinos have a mass and neutrino magnetic moment: 
\begin{equation}
\mu_{\upsilon}=\frac{3eG_{f}m_{\upsilon}}{8\pi^{2}\sqrt{2}}=\frac{3eG_{f}m_{e}m_{\upsilon}}{4\pi^{2}\sqrt{2}}\mu_{B}
\end{equation}
where $\mu_{B}$ is Bohr magneton. If the neutrinos have large magnetic moments, neutrinos transmissing through the Sun are effected by the solar magnetic field. Therefore, solar magnetic field can flip their spin and change left-handed neutrino to the right handed neutrino. Since right-handed neutrino is not detected by detector, it can be also responsible for the neutrino deficit. Okun, Voloshin and Vysotsky (OVV) [12] showed that the neutrino magnetic moment could be responsible for the deficiencies of solar neutrinos. Shortly after Akhmedov [13], Barbieri and Fiorentini [14] and Lim and Marciano [15] examined the combined effect of matter and magnetic field called Resonance Spin Flavor Precession (RSFP) and pointed out that it might lead to additional resonance besides the MSW resonance. After that for the solar neutrinos, RSFP effects was investigated in detail for chlorine and gallium experiments by Balantekin \textit{et al}.[16]. So far, several other studies related with RSFP have been studied in different aspects [17-22]. From the RSFP investigations of solar neutrinos, one can put a limit on $\mu B$, not the magnetic moment alone. In Ref. [23], authors examined the combined analysis of solar neutrinos and KamLAND data [24] and placed a limit on the $\mu B$. The limits on the magnetic moment come from astrophysical bounds, Supernova 1987A and solar neutrino experiments looking neutrino-electron scattering [25-28]. Detailed discussion on neutrino magnetic moment is given in [29].  

In this article, previous work [23] is extended with the new KamLAND data [30]. Altough detailed information of the solar magnetic field is required in the analysis of the RSFP framework, unfortunately, magnetic field profile in the Sun is not well known. Wood-Saxon shape of magnetic field profile is chosen here. RSFP formalism and analysis are given in the second section. Results and conclusion are presented in section 3.

\section{Formalism and Analysis}
In two generations case of Dirac neutrinos, the evolution equation for a neutrino passing through the matter and a magnetic field B is
\begin{equation}
i\frac{d}{dt}\left[
\begin{array}
[c]{c}%
\nu_{e_{L}}\\
\nu_{\mu_{L}}\\
\nu_{e_{R}}\\
\nu_{\mu_{R}}%
\end{array}
\right]  =\left[
\begin{array}
[c]{cccc}%
\frac{\delta m^{2}}{2E_{\nu}}sin^{2}\theta_{12} + V_{e} & \frac{\delta m^{2}}{4E_{\nu}}sin2\theta_{12} & \mu_{ee}B & \mu_{e\mu}B\\
\frac{\delta m^{2}}{4E_{\nu}}sin2\theta_{12} & \frac{\delta m^{2}}{2E_{\nu}}cos^{2}\theta_{12} + V_{\mu} & \mu_{{\mu}e}B & \mu_{\mu\mu}B\\
\mu^{*}_{ee}B & \mu^{*}_{{\mu}e}B & \frac{\delta m^{2}}{2E_{\nu}}sin^{2}\theta_{12} & \frac{\delta m^{2}}{4E_{\nu}}sin2\theta_{12}\\
\mu^{*}_{e\mu}B & \mu^{*}_{\mu\mu}B & \frac{\delta m^{2}}{4E_{\nu}}sin2\theta_{12} & \frac{\delta m^{2}}{2E_{\upsilon}}cos^{2}\theta_{12}%
\end{array}
\right]  \left[
\begin{array}
[c]{c}%
\nu_{e_{L}}\\
\nu_{\mu_{L}}\\
\nu_{e_{R}}\\
\nu_{\mu_{R}}%
\end{array}
\right]
\end{equation}
where $\theta_{12}$ is the vacuum mixing angle, $\delta m^{2}$ is the difference of the squares of the masses and $E_{\nu}$ is the neutrino energy. $V_{e}$ and $V_{\mu}$ are matter potentials for an unpolarized medium given as 

\begin{equation}
 V_{e}=\frac{G_{f}}{\sqrt{2}}(2N_{e}-N_{n}) \text{ \ \ \ \ \ \ }V_{\mu}=-\frac{G_{f}}{\sqrt{2}} N_{n}
\end{equation}
where $N_{e}$ and $N_{n}$ are electron and neutron number density respectively and $G_{f}$ is the Fermi constant.
In this analysis, results are found numerically via the diagonalization of the Hamiltonian in equation (2) which was discussed in detail in [16]. Altough, there are other various magnetic field profiles examined in the literature [17-20], magnetic field profile is taken to be Wood-Saxon shape of the form, as shown in figure 1:

\begin{equation}
B(r)=\frac{B_{0}}{1+\exp[10(r-R_{\odot})/R_{\odot}]}\label{3}%
\end{equation}
where $B_{0}$ is the strength of the magnetic field at the center of the Sun.
To calculate the best fits and confidence levels of allowed regions in the neutrino parameter space ($\delta m^{2}$ and $tan^{2}\theta_{12}$), common way in the literature is called $\chi^{2}$ analysis [31-34]. 'Covariance approach' is used to find the allowed regions. In this method, least-squares function for solar data is 
\begin{equation}
\chi_{_{\odot}}^{2}=\sum_{i_{1},i_{2}}^{N_{\exp}}(R_{i_{1}}^{(\exp)}-R_{i_{1}%
}^{(thr)})(V^{-1})_{i_{1}i_{2}}(R_{i_{2}}^{(\exp)}-R_{i_{2}}^{(thr)})\label{4}%
\end{equation}
where $V^{-1}$ is the inverse of the covariance matrix of experimental and
theoretical uncertainties, $R_{i}^{(\exp)}$ is event rate calculated in the
$i^{th}$ experiment and $R_{i}^{(thr)}$ is the theoretical event rate calculated for
$i^{th}$ experiment.For all solar neutrino experiments, chlorine (Homestake), gallium (SAGE, GALLEX, GNO), Super-Kamiokande and SNO, expressions of theoretical event rates are given in detail in [35]. 
Finally, one needs KamLAND data for the global analysis:
\begin{equation}
 \chi^{2}_{Global}= \chi_{_{\odot}}^{2}+\chi^{2}_{KamLAND}  
\end{equation}
 
\section{Results and Conclusions}

Allowed region of the neutrino parameter space for KamLAND data within the MSW framework alone is shown in figure 2 at 95\% CL. Joint analysis of solar neutrino and KamLAND data is given in Figure 3 at different $\mu B$ values at 95\% CL and projection of the global $\Delta\chi^{2}$ on $\mu B$ is shown in figure 4. One can see from figure 3 that as $\mu B$ values are increasing, the allowed regions in the LMA region are getting smaller and vanishes when $\mu B$ is greater than $1.2\times10^{-7}\mu_{B}G$ at 95\% CL. As shown in figure 4, the best minimum is at $\mu B=0.4\times10^{-7}\mu_{B}G$. One can find a limit on the $\mu B$ from the figure 4 for different confidence intervals. Such as: $\mu B<0.7\times10^{-7},1.0\times10^{-7},1.4\times10^{-7}\mu_{B}G$ for the $1\sigma$,  $2\sigma$, $3\sigma$ limits, respectively. Direct limits of neutrino magnetic moment from new experiments under study will be expected lower than $\mu <10^{-12}\mu_{B}$ or 1 order of magnitude lower [36-39]. To get such a limit, according to results found here, magnetic field B in the Sun must be higher than $10^{6} G$. However, since the limit on the magnetic field strength from helioseismological observations of the sound speed profile is about $10^{7}G$ [40] $\mu B$ found in this paper is too high to put such a lower limit on $\mu$ and one can say that RSFP scenerio does not have a crucial role on the solar neutrino data which agrees with the results of [22].

\bigskip

\bigskip

\textbf{Acknowledgments}

I would like to thank to A. Baha Balantekin for his suggestion of this research topic and for his kind help. I also thank to TUBITAK (The Scientific and Technological Research Council of Turkey) for BIDEB-2219 grant.

\pagebreak

\begin{figure}
[t]
\begin{center}
\includegraphics{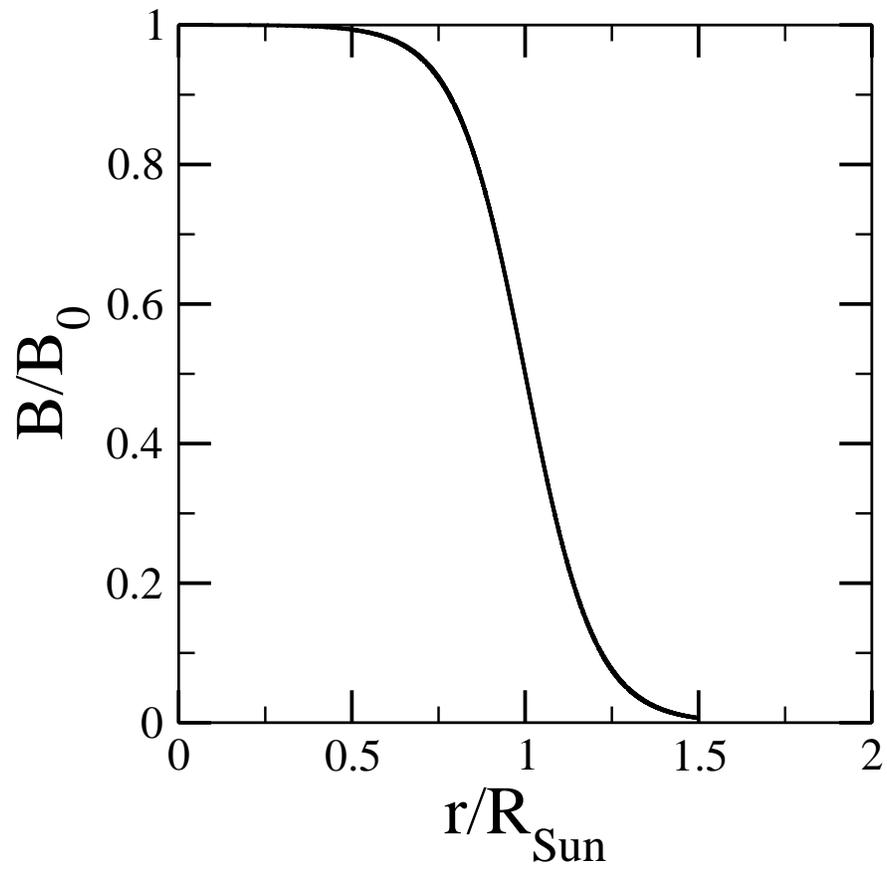}%
\caption{Magnetic field profile of Wood-Saxon shape.\label{fig1}}%
\end{center}
\end{figure}

\begin{figure}
[t]
\begin{center}
\includegraphics{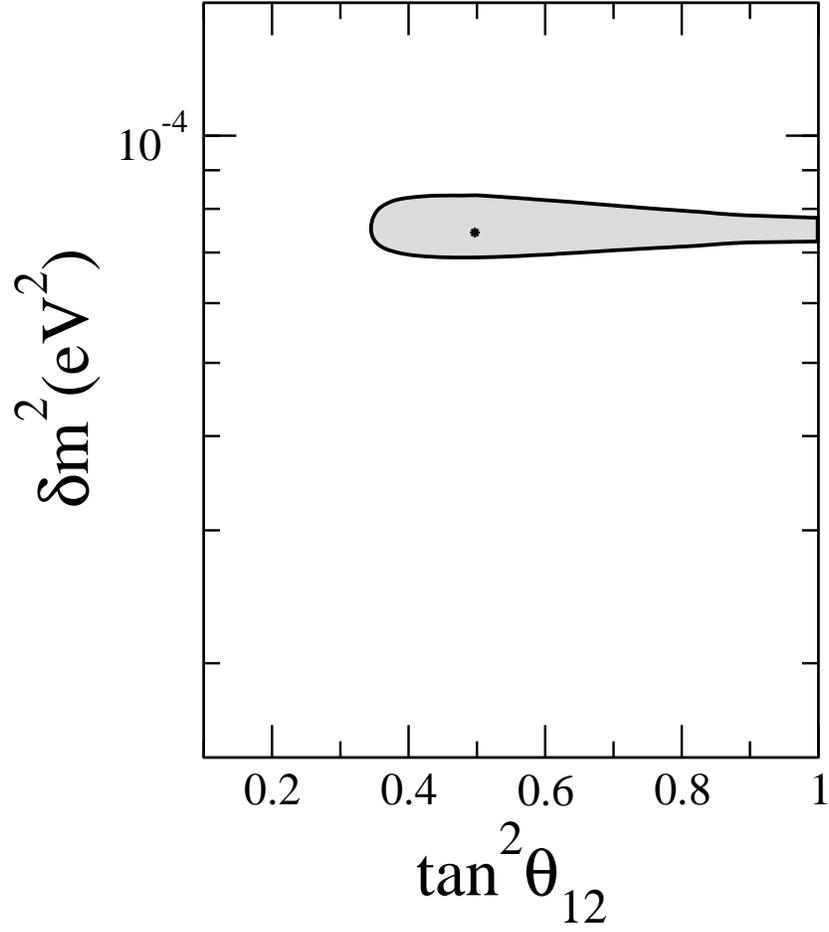}%
\caption{95\% confidence level interval allowed by KamLAND experiment within the MSW framework alone.\label{fig2}}%
\end{center}
\end{figure}

\begin{figure}
[t]
\begin{center}
\includegraphics{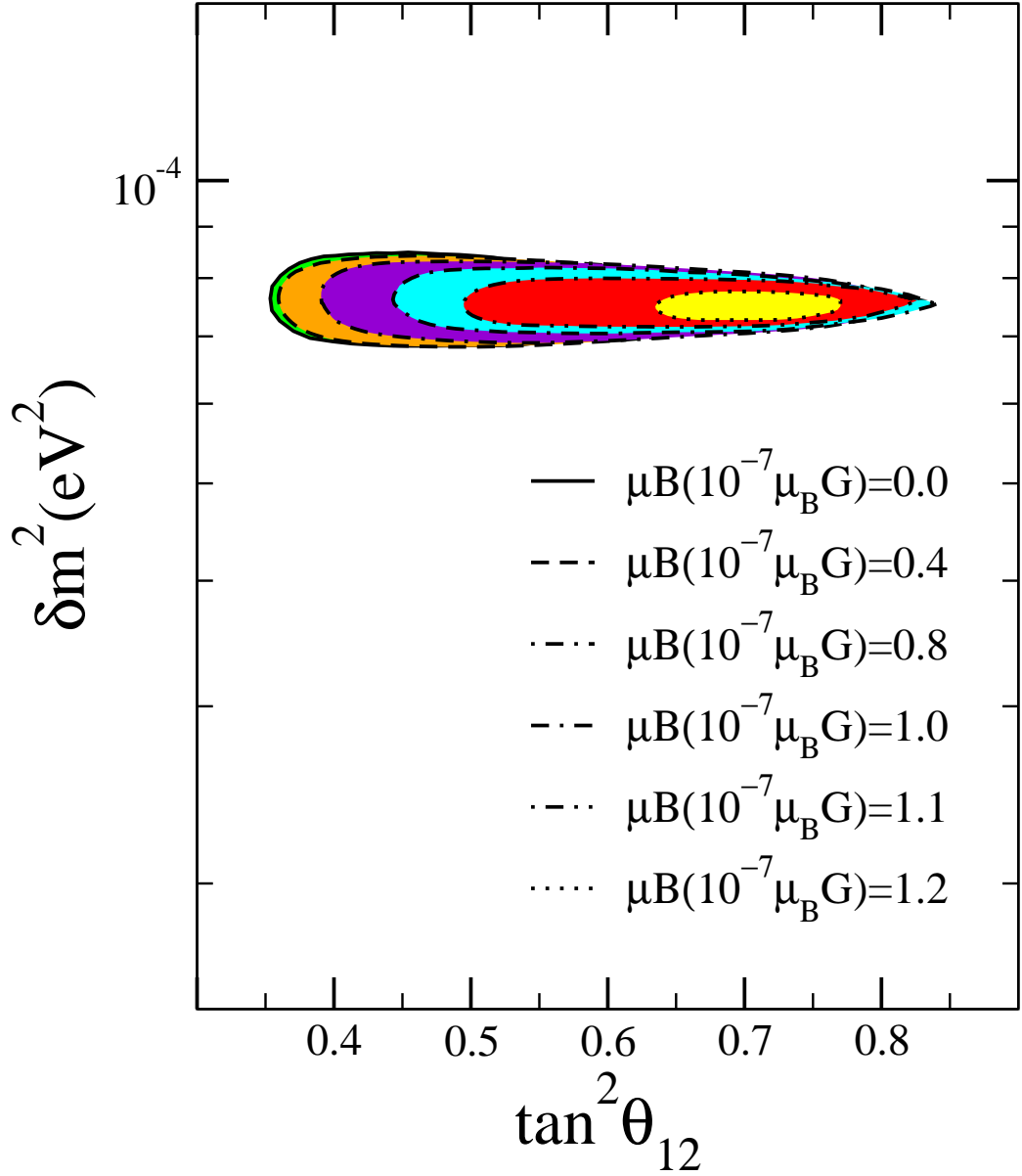}%
\caption{Three parameter 95\% CL intervals for the combine analysis of the solar and KamLAND data at some different $\mu B$ values.$\mu B(10^{-7}\mu_{B}G)=0.0,0.4,0.8,1.0,1.1,1.2$ from outside to inside.\label{fig3}}%
\end{center}
\end{figure}

\begin{figure}
[t]
\begin{center}
\includegraphics{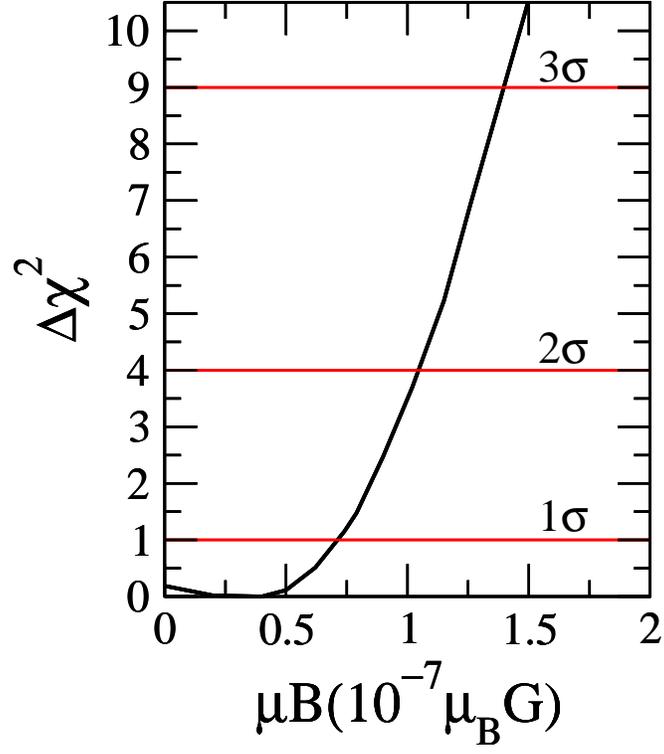}%
\caption{Projection of the global $\Delta\chi^{2}$ on $\mu B$.\label{fig4}}%
\end{center}
\end{figure}

\end{document}